\documentclass[aps,prb,twocolumn,superscriptaddress,showpacs]{revtex4}

\usepackage{graphicx}% Include figure files
\usepackage{dcolumn}% Align table columns on decimal point
\usepackage{bm}% bold math

\def\kdp{{\bf k}$\cdot${\bf p}}
\def\qsgw{\mbox{QS$GW$}}

\begin{document}
   
\preprint{APS/123-QED}

\title{Strain-Induced Conduction Band Spin Splitting in GaAs from First Principles Calculations}
\author{Athanasios N. Chantis} 
\affiliation{Theoretical Division, Los Alamos National Laboratory,
Los Alamos, New Mexico, 87545, USA}

\author{Manuel Cardona} 
\affiliation{Max Planck Institut f{\"u}r Festk{\"o}rperforschung, Heisenbergstrasse 1, D-70569 Stuttgart, Germany}

\author{Niels E. Christensen} 
\affiliation{Department of Physics and Astronomy, University of Aarhus, DK-8000 Aarhus C, Denmark}

\author{Darryl L. Smith} 
\affiliation{Theoretical Division, Los Alamos National Laboratory,
Los Alamos, New Mexico, 87545, USA}

\author{Mark van Schilfgaarde}
\affiliation{School of Materials, Arizona State University,
Tempe, Arizona, 85287-6006, USA}

\author{Takao Kotani}
\affiliation{School of Materials, Arizona State University,
Tempe, Arizona, 85287-6006, USA}

\author{Axel Svane} 
\affiliation{Department of Physics and Astronomy, University of Aarhus, DK-8000 Aarhus C, Denmark}

\author{Robert C. Albers} 
\affiliation{Theoretical Division, Los Alamos National Laboratory,
Los Alamos, New Mexico, 87545, USA}
%draft
\date{\today}

\begin{abstract}

We use a recently developed self-consistent $GW$ approximation to present
first principles calculations of the conduction band spin
splitting in GaAs under $[110]$ strain.  
The spin orbit
interaction is taken into account as a perturbation to the scalar
relativistic hamiltonian.  These are the first calculations of conduction
band spin splitting under deformation based on a quasiparticle approach; and because the
self-consistent $GW$ scheme accurately reproduces the relevant band
parameters, it is expected to be a reliable predictor of spin splittings.
We also discuss the spin relaxation time under $[110]$ strain and show that
it exhibits an in-plane anisotropy, which can be exploited to obtain the magnitude and \emph{sign}
of the conduction band spin splitting experimentally.  
\end{abstract}

\maketitle

\section{Introduction} 

The increasing prospect of utilizing spin electronics with conventional semiconductors, 
calls for quantitative predictions of the spin relaxation of electrons 
in these materials.~\cite{awschalom07,kato03}
In semiconductors without inversion symmetry, the spin relaxation rate is related to the relativistic splitting in 
the conduction band, an effect which also induces spin precession and is relevant for spin
transport and injection. \cite{rats84,arc85,ar92,sat05} 
In zinc-blende semiconductors, which are the most promising for spintronic applications, it is
widely accepted that D'yakonov-Perel' (DP) \cite{dp71} is the dominant spin relaxation mechanism.
Generally speaking, this mechanism is present when the spin degeneracy of the conduction band is lifted.
The spin splitting can be viewed as a {\bf k}-dependent effective magnetic 
field $\mathbf{\Omega}(\mathbf{k})$ which under certain
scattering conditions relaxes the average spin of the ensemble. 
The strength of the effective field depends on the material.
In the general form we can add this field to
the Hamiltonian as an effective Zeeman term, $H(\mathbf{k})= 1/2 \mathbf{\sigma}\cdot \mathbf{\Omega}$,
where $\mathbf{\Omega}(\mathbf{k})$ (with the dimensions of energy) is proportional to ${\mathbf{B_{eff}}(\mathbf{k})}$.
In the zinc-blende crystal structure there is no inversion symmetry; this leads  
to an $\mathbf{\Omega}$ field with components~\cite{gd55,rash61}
\begin{equation}
\Omega_{\rm D}^{i}=2\gamma k_{i}
(k^{2}_{i+1}-k^{2}_{i+2})
\label{eq:dress}
\end{equation} 
where $i=x,i+1=y,i+2=z$, the indices obey a cyclic relationship ($i+3=1$), and $\gamma$ is a
constant that depends on the bulk properties of the material. This effective field was first introduced 
by Dresselhaus~\cite{gd55}.
In uniaxially deformed crystals there is an additional effective field~\cite{dmpt86,pikus,birpikus62,safarov83}
\begin{eqnarray}
\Omega^{i}_{stress} = C ( \epsilon_{i,i+1} k_{i+1} - \epsilon_{i,i+2} k_{i+2} )+\nonumber\\
B k_i ( \epsilon_{i-2,i-2} - \epsilon_{i-1,i-1} )
\label{eq:stress}
\end{eqnarray}
$\epsilon_{ij}$ is the strain tensor, $C$ and $B$ are material-dependent constants.
The first part of the effective field originates from off-diagonal components of the strain tensor
while the second part originates from diagonal ones. The second part appears only due to
the spin orbit mixing of $p$ and $d$ states and therefore should be much
weaker than the first part~\cite{cmt84}, but a numerical estimation of $B$ has not yet been performed, probably because of uncertainties concerning the $\Gamma_{12}$ states. 

The data for the values of $\gamma$, $C$ and $B$ in various zinc-blende 
semiconductors are very sparse. The most studied case is GaAs. However, the
experimental data for $\gamma$ show a wide range of values, between 11.0 and 34.5~eV$\cdot${\AA}$^3$.~\cite{mynote}
Theoretical calculations of this parameter also show a wide range of predicted values.
Calculations based on \kdp~method predict a value between 25 and 30 eV$\cdot${\AA}$^3$.~\cite{mynote}
However, a first principles calculation by Cardona \emph{et al} \cite{ccf88} predicted a value of
15 eV$\cdot${\AA}$^3$. Our recent first principles calculation~\cite{chantis06} predicted 
a value of 8.5 eV$\cdot${\AA}$^3$,
a lot smaller than the commonly cited value of 27.5 eV$\cdot${\AA}$^3$. Recently, Krich \emph{et al}~\cite{kirch}
used a semi-classical approach to estimate the effect of  $\Omega_{\rm D}$ on
the mean and variance of the conductance in closed quantum dots and compared the results of their 
model with the experiment in Ref.~\onlinecite{zumbuhl}. They were able to reach 
a good agreement only when they used our value of $\gamma$, suggesting that the value of
$\gamma$ in GaAs must be around 9~eV$\cdot${\AA}$^3$.   

Also our current knowledge of $C$ and $B$ in GaAs is far from being satisfactory. 
The magnitude of $C$ was estimated experimentally in Refs.~\onlinecite{beck} and ~\onlinecite{gmt}
to be 8.1$\pm$2.5 eV$\cdot${\AA} and 3.9  eV$\cdot${\AA}, respectively. The values of $C$
calculated with Linear Combination of Atomic Orbitals (LCAO) and pseudopotentials are 3.75 
and 11.24 eV$\cdot${\AA}, respectively~\cite{ccf88}. 
The calculations are able to define also the sign of $C$; 
in Ref.~\onlinecite{ccf88} it was found to be opposite to that of $\gamma$.~\cite{ccfnote}
No other attempt has been made since to calculate the sign of $C$. 
To our knowledge the magnitude and sign of $B$
has not been estimated experimentally nor theoretically so far. Although,
M. I. D'yakonov \emph{et al}~\cite{dmpt86} showed that the spin
relaxation time has a very weak dependence on applied pressure 
upon application of [100] strain, possibly indicating that the magnitude of $B$ 
is negligibly small.

In this work we use a recently developed {\it ab inito} method 
based on the $GW$ approximation to predict
these parameters for GaAs. 
We will try to answer the important questions about the strengths and 
signs of the spin splittings caused by the two mechanisms, Eqs. \ref{eq:dress} and \ref{eq:stress}. 

\section{Method} 
The $GW$ approximation can 
be viewed as the first term in the expansion 
of the non-local energy dependent self-energy $\Sigma(\bf{r},\bf{r},\omega)$
in the screened Coulomb interaction $W$. From a more physical point of view
it can be interpreted as a dynamically screened Hartree-Fock approximation plus
a Coulomb hole contribution \cite{Hedin}. 
It is also a prescription for mapping the non-interacting Green function to the dressed
one, $G^0 \to G$. In the Quasiparticle Self-consistent $GW$ (\qsgw) method a prescription is given on how to map
$G$ to a new non-interacting Green function $G \to G^0$. 
This is used for the input to a new iteration; we repeat the procedure 
$G^0 \to G \to G^0 \to ....$ until converegence is reached. 
Thus \qsgw\ is a self-consistent perturbation theory, where the
self-consistency condition is constructed to minimize the size of the
perturbation.  \qsgw\ is parameter-free, independent of basis set and of
the LDA \cite{kotani}.  
The method is described in great detail in Refs.~\onlinecite{kotani} and ~\onlinecite{mark06}.
It has been shown that \qsgw\ reliably describes the band structure in a wide
range of materials  \cite{mark,faleev,chantis06,chantis07}.

\begin{figure}[t]
\includegraphics[angle=0,width=0.3\textwidth,clip]{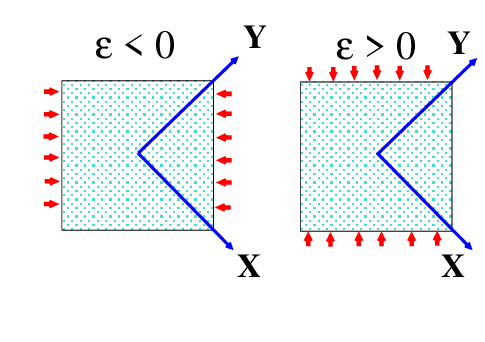}% The 2 deformation directions
\caption{\label{fig:split3def} The left panel represents the applied deformation for $\epsilon<0$ 
and the right for $\epsilon>0$ .}
\end{figure}

The \qsgw\ method in the current implementation uses the 
Full Potential Linear Muffin Tin Orbital (FP-LMTO) 
method \cite{OKA75,markbook},
so we make no approximations for the shape of the crystal potential.
The smoothed LMTO basis includes orbitals 
with $l \le l_{max}=5$ and both 3$d$ and 4$d$  are included
in the basis. $4d$ are added in the form of local orbitals \cite{mark06}--an orbital 
strictly confined to the augmentation sphere, which has no envelope function at all.
%$7p$ are added as a kind of extended 'local orbitals' the 'head' of which is evaluated at
%an energy far above Fermi level \cite{MarkPRB06}, instead of making the
%orbital vanish at the augmentation radius a smooth Hankel'tail' is attached to the orbital.
As \qsgw\ gives the self-consistent 
solution at the scalar relativistic level, 
we add the spin-orbit operator, $H_{SO}$, 
%=\mathbf{L}\cdot\mathbf{S}/2c^{2}$ 
as a perturbation (it is not included in the self-consistency cycle). 

It has also been shown that the \qsgw\ method systematically overestimates the fundamental
band gap in semiconductors by an amount of a few tenths of an eV, independent of the
magnitude of the gap~\cite{mark}. 
This error is related to the fact that the 
vertex correction is not taken into account in the method and when taken into account
a nearly perfect agreement with experiment is achieved~\cite{shishkin}. 
Here, in order to obtain highly accurate results with less computational effort, 
we take a simple but somewhat heuristic approach to correct the error.
We considered a `hybridized' \qsgw+LDA Hamiltonian with
\begin{eqnarray}
H_\alpha +  H_{\rm SO} = H_{\rm LDA} + (1-\alpha)(\widetilde{\Sigma}-
V^{\rm LDA}_{\rm xc}) + H_{\rm SO}.
\end{eqnarray}
%(Note that $H_{\rm LDA}$ and $V^{\rm LDA}_{\rm xc}$ are also determined
%from the self-consistent density given by $H_0$). 
In Ref.~\onlinecite{chantis06} we found
that for all III-V and II-VI semiconductors studied a value of $\alpha \approx 0.2$ gives
excellent agreement of calculated band gap and other important band parameters with
experiment. 

% ----------- Table I ---------------

\begin{table}
\caption{\label{tab:table1} Important band parameters for GaAs.  $E_{0}=E(\Gamma^c_6)-E(\Gamma^v_8)$ and $E^{\prime}_{0}=E(\Gamma^c_7)-E(\Gamma^v_8)$ are the energies of the
  first two conduction bands at the $\Gamma$-point.  $\Delta_{\rm{}SO}$ and $\Delta^{\prime}_{\rm{}SO}$ are the spin-orbit splittings between $\Gamma_{8}$
  and $\Gamma_7$ for valence and conduction bands, respectively.
$m^{\Gamma}_{c}/m$ is the conduction band effective mass at
  $\Gamma$.  Energies are in eV; $\gamma$ is in eV$\cdot\rm{}\AA^{3}$, $C$ and $B$
  are in eV$\cdot\rm{}\AA$. An asterisk in front of the value indicates that this is a calculated value from
another theoretical method.
 }
%The sign of $\Delta^{-}$ is inferred from the sign of the off-diagonal matrix elements of $H_{{\rm SO}}$ in the eigenfunction basis.
\begin{ruledtabular}

\begin{tabular}{lccc}
                               &QSGW+LDA&    QSGW   & Expt        \\
\hline
$E_{0}$                        &   1.52   &   1.80     &   1.52\footnotemark[1]  \\
\hline
$E^{\prime}_{0}-E_{0}$        &   2.89   &   2.81     &   3.08\footnotemark[1]   \\
\hline
$\Delta_{{\rm SO}}$            &   0.336   &         &   0.341\footnotemark[1]   \\
\hline
$\Delta^{\prime}_{{\rm SO}}$ &  0.174   &        &     \\
%\hline
%$\Delta^{-}$                 &  -0.12    &        &    \\
\hline
$m^{\Gamma}_{c}/m$           &   0.069   &   0.076     &   0.067\footnotemark[1]  \\
\hline
$\gamma$                 &  +8.5   &  +6.4     &  11.0-34.5\footnotemark[2]  \\
\hline 
$C$                     &  +6.81     &   +5.39    &  3.9\footnotemark[5], 4.0\footnotemark[6], 
5.3\footnotemark[3], 8.1$\pm$2.5\footnotemark[4],\\
& & &*-3.74\footnotemark[7],*-11.2\footnotemark[8],*2.0\footnotemark[9],\\
& & &  *5.0\footnotemark[10],*4.9\footnotemark[11]\\
\hline
$B$          & +2.13   &  +1.7     &      \\
\end{tabular}
\end{ruledtabular}
\footnotetext[1]{From Ref.~\protect\onlinecite{madelung}}
\footnotetext[2]{From Ref.~\protect\onlinecite{mynote}}
\footnotetext[3]{From Ref.~\protect\onlinecite{gmt,pikus}; where only the absolute value is reported}
\footnotetext[4]{From Ref.~\protect\onlinecite{beck}; where only the absolute value is reported}
\footnotetext[5]{From Table IX of Ref.~\protect\onlinecite{ccf88}; where only the absolute value is reported}
\footnotetext[6]{From Ref.~\protect\onlinecite{cmt84}; where only the absolute value is reported}
\footnotetext[7]{Calculated with Pseudopotentials; From Table IX of Ref.~\protect\onlinecite{ccf88}}
\footnotetext[8]{Calculated with LCAO; From Table IX of Ref.~\protect\onlinecite{ccf88}}
\footnotetext[9]{Calculated with Pseudopotentials; From Ref.~\protect\onlinecite{cmt84}}
\footnotetext[10]{Calculated with LCAO; From Ref.~\protect\onlinecite{cmt84}}
\footnotetext[11]{Calculated with the three-band \kdp~method; From Ref.~\protect\onlinecite{pikus}}
\end{table}

All band parameters presented in Table I, \emph{except} $C$ and $B$, are calculated 
for the undistorted lattice structure. All signs are presented with the convention
that the \emph{anion is at the origin and cation at (0.25,0.25,0.25)}.
We see that overall the \qsgw\ is in good agreement with experiment but the  
`hybridized' \qsgw+LDA Hamiltonian is in even better agreement.
For $C$ and $B$ we calculate self-consistently the self-energy
and charge density under the corresponding deformation. We found that
if instead we use the self-consistent self energy of the undistorted structure
the value of $C$ differs from that presented in Table I by $\approx$5$\%$.
  In these calculations
the atomic positions were allowed to relax within LDA in order to account for the
displacement of the anion and cation sublattices relative to each other.
For a pure shear deformation in the [111] direction this displacement can be
viewed as a length change $\Delta l$ of the [111] bond described by the internal 
strain parameter $\zeta$, $\Delta l=3(1-\zeta)\epsilon/\sqrt{4}$. 
Our calculated value of $\zeta$ is 0.53, in good agreement with previous calculations
~\cite{rcc99,bpc84}  

We apply two different deformations, the first of which is described by the following strain tensor:
\begin{equation}
{\bf\varepsilon}=\left(
\begin{array}{ccc}
\epsilon_1& \epsilon& 0\\
\epsilon& \epsilon_1& 0\\
0& 0&\epsilon_2 
\end{array}\right)\;
\label{eq:def1}
\end{equation}
where $\epsilon_1=$0.0025186, $\epsilon_2=$-0.0049628 
and $\epsilon=$0.0074814.
This tensor conserves the volume and
induces the $B$ related term in $\Omega^{i}_{stress}$.
To separate this term from the $C$ related term we also
performed a calculation with a deformation described by the following strain tensor:
\begin{equation}
\mathbf{\varepsilon}=\left(
\begin{array}{ccc}
0& \epsilon    & 0 \\
\epsilon& 0& 0\\
0& 0&0 
\end{array}\right)\;
\label{eq:def2}
\end{equation}
This strain tensor conserves the volume only to first order of deformation
but only the first term ($C$ related) is present in $\Omega^{i}_{stress}$.

\section{results} 
\subsection{Spin splittings}
\begin{figure}[tbp]
\includegraphics[angle=0,width=0.5\textwidth,clip]{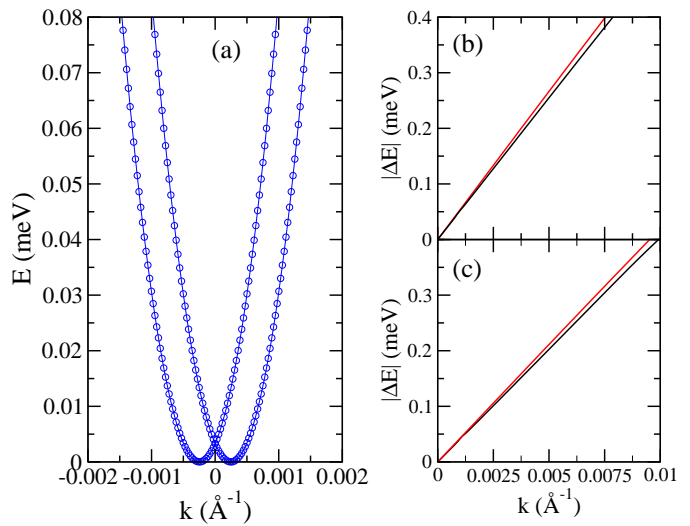}
\caption{
(a) The shift of the conduction band minimum away from the $\Gamma$ point in GaAs under deformation
given by Eq. \ref{eq:def1} ('hybridized' method). (b) The magnitude of the conduction
band splitting along the [010] direction for the case of the 'hybridized' Hamiltonian, red line with
deformation (\ref{eq:def1}) black line with deformation (\ref{eq:def2}) (c) same as (b) but 
for \qsgw\ Hamiltonian. }
\label{fig:fig1}
\end{figure}
In Fig.~\ref{fig:fig1} 
we show the $k$-dependence of the conduction band splitting along [010]
for the case of 'hybridized' and \qsgw\ calculations.
Figure~\ref{fig:fig1}(a) shows the energy dispersion around the $\Gamma$ point for the
case of the 'hybridized' Hamiltonian and the deformation given by Eq.~\ref{eq:def1}.
Along the [010] direction $\Omega_{\rm D}^{i}$ vanishes and the dispersion is given
by 
\begin{equation}
E_{\pm}(k)=\hbar^2k^2/2m_{eff} \pm 1/2 \left|A\right| k 
\end{equation}
where
\begin{equation}
\left|A\right|=\sqrt{(C\epsilon)^2+B^2(\epsilon_2-\epsilon_1)^2}
\end{equation}
and the strain components were introduced in Sect. II. 
%$\epsilon$=0.0074814, $\epsilon_1$=0.0025186 
%and $\epsilon_2$=-0.0049628.
%here, $\epsilon$=0.0074814, $\epsilon_1$=0.0025186 and $\epsilon_2$=-0.0049628.
Correspondingly, the conduction band minimum shifts to 
\begin{equation}
k_{\pm}=\pm m_{eff}/2\hbar^2 \left|A\right|
\end{equation}
In Figs.~\ref{fig:fig1}(b) and (c) we show the magnitude of the splitting along
the [010] direction for the case of 'hybridized' and \qsgw\ Hamiltonians, respectively.
The black solid lines are for strain (\ref{eq:def1}) and the red solid lines
for (\ref{eq:def2}). The slope of the red line is equal to
\begin{equation} 
s_1=\sqrt{(C\epsilon)^2+B^2(\epsilon_2-\epsilon_1)^2} 
\end{equation}
while the slope of the black line is equal to 
\begin{equation}
s_2=\left|C\right|\left|\epsilon\right|
%s_2=C\left|\epsilon\right|
\end{equation} 
Hence 
\begin{equation}
\left|C\right|=s_2/\left|\epsilon\right|
%C=s_s/\left|\epsilon\right|
\end{equation} 
and
\begin{equation} 
\left|B\right|=\sqrt{s_1^2-s_2^2}/\left|\epsilon_2-\epsilon_1\right|
%B=\sqrt{s_1^2-s_2^2}/\left|\epsilon_2-\epsilon_1\right|
\end{equation}
The magnitudes of $C$ and $B$ extracted with this procedure are given in Table I.
As expected, in the case of the 'hybridized' method they are slightly larger than
in the \qsgw\ because the 'hybridized' band structure has a smaller band gap.   
However, the ratio of $C/B$ remains nearly constant; it is equal to 3.197 in the 'hybridized'
method and 3.171 in \qsgw. The magnitude of $C$ is in good agreement 
with experiments in both the \qsgw\ and the 'hybridized' methods but the latter should be
trusted more due to better agreement of the other band parameters with experimental values. 
We also note that in the experimental determination of $C$, the $B$ terms were considered to be negligible.
 This corresponds to extracting the value of $C$ directly from $s_1$; according to
our calculations this would yield an inaccurate value of $C=7.14$ eV$\cdot$$\AA$ with an error of 5$\%$, much
less than the experimental error in Ref.~\cite{beck}.
\begin{figure}[tbp]
\includegraphics[angle=0,width=0.38\textwidth,clip]{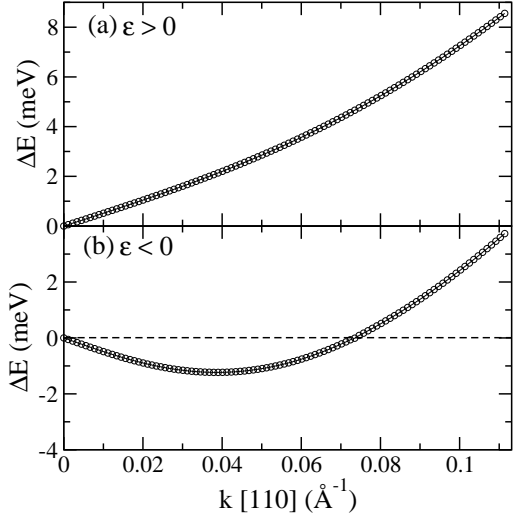}
\caption{
(a) The conduction band splitting along the [110] direction for the case of 'hybridized' 
Hamiltonian, applied strain (\ref{eq:def2}) for $\epsilon>0$ (c) same as (a) but for $\epsilon<0$. }
\label{fig:fig2}
\end{figure}
\begin{figure}[htbp]
\includegraphics[angle=0,width=0.38\textwidth,clip]{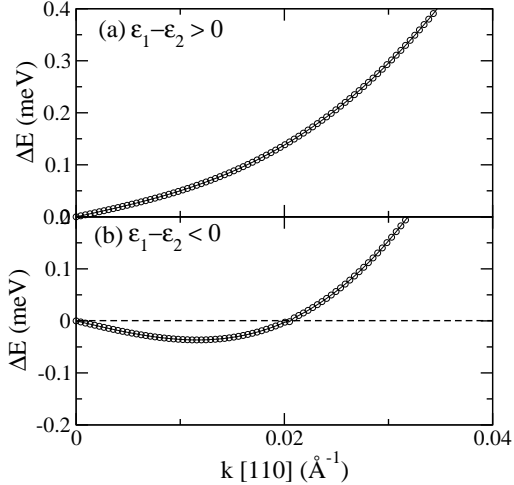}
\caption{
(a) The conduction band splitting along the [110] direction for the case of 'hybridized' Hamiltonian, 
under deformation (\ref{eq:def3}) with $(\epsilon_1-\epsilon_2)>0$ (b) same as 
(a) but for $(\epsilon_1-\epsilon_2)<0$. }
\label{fig:fig3}
\end{figure} 
The $\Omega_{\rm D}^{i}$ is highly anisotropic and completely vanishes in certain directions but
is present for a general direction. It is therefore interesting to compare 
$\Omega_{\rm D}^{i}$ with $\Omega^{i}_{stress}$. We start by comparing $C$ to $\gamma$.
When we apply deformation (\ref{eq:def2}), the dispersion along the [110] 
direction is 
\begin{equation}
E(k)=\hbar^2k^2/2m_{eff} \pm 1/4\left(\gamma k^3+ C\epsilon k\right)
\end{equation}
Thus the Dresselhaus and the stress terms can either add or subtract, depending
on the relative sign of $C\epsilon$ and $\gamma$. 
If they subtract the splitting will be zero at $k_{rev}=\sqrt{C/\gamma}\sqrt{\epsilon}$
%and for bigger $k$ along (110) it will change sign.
i.e. the spin splitting reverses its sign at $k_{rev}$.
In Table I it is seen that $\sqrt{C/\gamma}=0.895 \AA^{-1}$, which means that for
the deformation (\ref{eq:def2}) the spin splitting along [110] will change sign
at $k_{rev}=0.077$~{\AA}$^{-1}$. 
%This is way out of the range shown in Fig.~\ref{fig:fig1},
%a small correction due to nonlinearity may apply.
%However, applied experimental deformations are orders of manitude less than
%what is shown in ~\ref{eq:def2}. A more likely value for $\epsilon$ would be
%between 0.0005 and 0.00005, which would result in $k_{rev}$ approximately 
%between 0.02 and 0.006 $\AA^{-1}$.  
As shown in Fig. 1, $\epsilon$
can be either negative or positive,
therefore such cancellation will always occur depending on the sign of the $C\cdot\epsilon\cdot k$ product. 
For example, if $k>0$, whether it occurs for $\epsilon>0$ or $\epsilon<0$ depends on the sign of $C$.
In Ref.~\onlinecite{chantis06} we determined the sign of $\gamma$ according to 
conventions in Ref.~\onlinecite{ccf88}. Here we will determine the sign of $C$ relative
to the sign of $\gamma$ by simply plotting $\Delta E$ along [110] for
positive and negative $\epsilon$. In Fig.~\ref{fig:fig2} we show such plot for
the deformation (\ref{eq:def2}) with positive and negative $\epsilon$. 
The splitting is linear only in the vicinity of the $\Gamma$ point, 
away from the $\Gamma$ point the cubic term is clearly visible. For $\epsilon>0$ the 
two contributions add (Fig.~\ref{fig:fig2}(a)) but for $\epsilon<0$ they oppose 
each other (Fig.~\ref{fig:fig2}(b)); for $k>0.074 \AA^{-1}$ the cubic term 
dominates and the splitting becomes positive. It is clear that $C$ and  $\gamma$ 
have the \emph{same sign} 
(according to the convention used here they are both positive).
The sign of B is defined in a similar way. We apply the deformation:
\begin{equation}
\mathbf{\epsilon}=\left(
\begin{array}{ccc}
\epsilon_1& 0    & 0 \\
0& \epsilon_1    & 0\\
0& 0    & \epsilon_2 
\end{array}\right)\;
\label{eq:def3}
\end{equation}
So that $\epsilon_2=-2\epsilon_1$. Then the splitting along the [110] direction is
$\Delta E \propto \left[\gamma k^3 + (\epsilon_1-\epsilon_2) B k\right]$. 
if $\Delta E$ crosses zero for $(\epsilon_1-\epsilon_2)>0$ then $B$ and $\gamma$ have 
opposite signs, otherwise $B$ and $\gamma$ have the same sign. 
As can be seen in Fig.~\ref{fig:fig3} we find that $B>0$.
\subsection{Spin relaxation}
After having reliably determined the values of material parameters that 
dictate the spin relaxation rate in GaAs it will be interesting to estimate
the spin relaxation time for a deformation like the one given by Eq.~\ref{eq:def1}.
In the Appendix we have derived the average spin relaxation time when
both $\mathbf{\Omega_{\rm D}}$ and $\mathbf{\Omega_{stress}}$ are present
\begin{widetext}
\begin{equation}
\frac{1}{\tau_{\perp,1}}=\gamma_l\left[\left[\frac{1}{3}\left(C\epsilon\right)^2+\left(B\Delta \epsilon\right)^2\right]\frac{m^2}{\hbar^2}\langle v^2 \tau_p(E)\rangle
+\frac{32}{105}\gamma^2\frac{m^6}{\hbar^6}\langle v^6 \tau_p(E)\rangle\right]+\frac{2}{3}\gamma_l B C \epsilon \Delta\epsilon \frac{m^2}{\hbar^2}\langle v^2 \tau_p(E)\rangle
\label{eq:sr1}
\end{equation}
\begin{equation}
\frac{1}{\tau_{\perp,2}}=\gamma_l\left[\frac{1}{3}\left[\left(C\epsilon\right)^2+\left(B\Delta \epsilon\right)^2\right]\frac{m^2}{\hbar^2}\langle v^2 \tau_p(E)\rangle
+\frac{32}{105}\gamma^2\frac{m^6}{\hbar^6}\langle v^6 \tau_p(E)\rangle\right]-\frac{2}{3}\gamma_l B C \epsilon \Delta\epsilon \frac{m^2}{\hbar^2}\langle v^2 \tau_p(E)\rangle
\label{eq:sr2}
\end{equation}
\begin{equation}
\frac{1}{\tau_{\parallel}}=\gamma_l\left[\frac{2}{3}\left[\left(C\epsilon\right)^2+\left(B\Delta \epsilon\right)^2\right]\frac{m^2}{\hbar^2}\langle v^2 \tau_p(E)\rangle
+\frac{32}{105}\gamma^2\frac{m^6}{\hbar^6}\langle v^6 \tau_p(E)\rangle\right]
\label{eq:sr3}
\end{equation}
\end{widetext}
Where $\parallel$ denotes axis parallel to the vector ${\bf N}=(0,0,1)$ and 
$\perp$ perpendicular to it. Namely, $(\perp,1)$ and $(\perp,2)$ are the axes along the [1$\bar{1}$0] and 
[110] crystal directions, respectively.
We see that there is an in-plane
anisotropy induced by the simultaneous presence of $\epsilon$
and $\Delta\epsilon=\epsilon_1 - \epsilon_2$ strain components (a similar anisotropy was observed for the circular piezo-birefringence and 
confinement-induced circular birefringence in GaAs~\cite{ksc98}).   
We can write the strain tensor (\ref{eq:def3}) in these axes: $\varepsilon=\epsilon_{i,j}\delta_{ij}$ with 
$\epsilon_{1,1}=\epsilon_{[110]}=\epsilon_1+\epsilon$, 
$\epsilon_{2,2}=\epsilon_{[1\bar{1}0]}=\epsilon_1-\epsilon$ and $\epsilon_{3,3}=\epsilon_{[001]}=\epsilon_2$. 
If we apply a uniaxial pressure (stress), $p$,  along [110] in this system of coordinates, then using the compliance constants $S_{11}$, $S_{12}$ and $S_{44}$ 
%from the stress-strain relationship we can see that 
we find $\epsilon_{[110]}=\left(S_{11}+S_{12}+S_{44}/2\right)p/2$, $\epsilon_{[1\bar{1}0]}=\left(S_{11}+S_{12}-S_{44}/2\right)p/2$ and $\epsilon_{[001]}=2S_{12}p/2$, hence, 
$\Delta\epsilon=\left(S_{11}-S_{12}\right)p/2$ and $\epsilon=(1/2)S_{44}p/2$. 
%The $\alpha_1$ and $\alpha_2$ coefficients depend on elasctic constants
%in the following way: $\alpha_1=1/3\left(1/(3G)+1/\mu\right)$ and $\alpha_2=1/3\left(1/(2\mu) - 1/(3G)\right)$, 
%where $G$ is the bulk modulus and 
%$\mu$ is the shear modulus.
According to Eqs. \ref{eq:sr1} and \ref{eq:sr2} the difference between 
the inverse of the spin relaxation time along [1$\bar{1}$0] and [110] is equal to 
\begin{eqnarray}
&&\frac{1}{\tau_{\perp,1}}-\frac{1}{\tau_{\perp,2}}=\nonumber\\
&&\frac{BC}{6}\gamma_l\frac{m^2}{\hbar^2}\langle v^2 \tau_p(E)\rangle S_{44}\left(S_{11}-S_{12}\right)p^2
\label{eq:anisotropy}
\end{eqnarray}
Hence, an experimental
setup similar to the one described above should be able to measure a linear increase of
the difference (\ref{eq:anisotropy}) with the square of applied pressure.  
The rate of increase should be proportional
to $B C$. In the experiment of Ref.~\onlinecite{dmpt86} the authors measured the increase of
the spin relaxation time with applied pressure along [100]. Such strain will only induce the
$B$ terms in Eqs. (\ref{eq:sr1})-(\ref{eq:sr3}). If we assume that the
applied strain is large enough to ignore the Dresselhaus term, then the spin
relaxation time should be isotropic and should increase linearly with the square of applied strain. 
However, unlike the experimental setup proposed here the
rate of increase is proportional to $B^2$. 
The experiment proposed here is independent of Dresselhaus terms no matter how small is the deformation, 
also the linear increase is proportional to $C B$ instead of $B^2$, hence it may be easier to detect.
Provided that the orientation of the As-Ga bond has been previousely determined, this
experiment can be used to find the sign of $B$ relative to that of $C$ from the sign of the difference (\ref{eq:anisotropy}). 
 
\section{conclusion}

We have presented first principles calculations of the magnitude and sign of bulk constants
that govern the DP spin scattering in GaAs under strain. To our knowledge, this was the first 
estimation of magnitude and sign of $B$. We find that both $C$ and $B$ have the
same sign as $\gamma$. Our value of $C$ is in good agreement with experiments. 
We have derived an expression for the spin relaxation time of electrons under
a strain given by Eq. \ref{eq:def1} and showed that the in-plane spin relaxation is anisotropic
in this case. We proposed an experiment that can exploit this anisotropy to deduce 
the magnitude and sign of $B$.  
%We presented an argument that spin scattering rate induced by very small strains can be of the
%same order of magnitude with that induced by the Dresselhaus term. This may have
%contributed to the big scatter of experimental values for $\gamma$ and should
%be taken into account whenever a connection between spin scattering rates and 
%$C$, $B$, $\gamma$ is made.

\begin{acknowledgments}

The work at Los Alamos was supported by DOE Office of Basic Energy Sciences Work Proposal Number 08SCPE973.

\end{acknowledgments}

\section{Appendix: Spin scattering rate} 

The momentum dependent spin relaxation time tensor is defined as:

\begin{equation}
\frac{1}{\tau_{i,i}(k)}=\gamma_l \tau_p(E) \left(\overline{\Omega^2}-\overline{\Omega^2_i}\right)
\end{equation}
and
\begin{equation}
\frac{1}{\tau_{i,j}(k)}=\gamma_l \tau_p(E) \left(\overline{\Omega_j\Omega_j}\right) \mbox{      $(i\neq j)$}
\end{equation}
Here $i={x,y,z}$ and the overbar denotes averaging over all directions of $\mathbf{k}$.
$\tau_p(E)$ is the momentum scattering time for an electron with energy $E$ and 
\begin{equation}
\gamma_l=\frac{\int^{+1}_{-1}\sigma(\cos\theta)(1-P_l(\cos\theta)
)d\cos\theta}{\int^{+1}_{-1}\sigma(\cos\theta)(1-\cos\theta)d\cos\theta}
\end{equation}
where $\sigma(cos\theta)$ is the electron scattering cross section and $P_l$ 
the Legendre polynomials. Here it is 
assumed that the electron scattering is elastic, the electron energy spectrum is
isotropic and the scattering cross section $\sigma(\mathbf{k},\mathbf{k^{\prime}})$
depends only on the scattering angle $\theta$.
$\mathbf{\Omega}$ is the total effective field and in 
our case we can write:
\begin{equation}
\mathbf{\Omega}=\mathbf{\Omega_{\rm D}}+\mathbf{\Omega_{stress}}
\end{equation} 
with the components as given in Eqs. (\ref{eq:dress}) and 
(\ref{eq:stress}). We apply a strain like that of tensor Eq. 
(\ref{eq:def1}) with the constraint
$\epsilon_2=-2\epsilon_1$, so as to conserve the volume. Then
%\begin{equation}
%\Omega_{\rm D}^{i}=2\gamma k_{i}
%(k^{2}_{i+1}-k^{2}_{i+2})
%\label{eq:dress}
%\end{equation} 
%and 
%\begin{eqnarray}
%\Omega^{i}_{stress} = C ( \epsilon_{i,i+1} k_{i+1} - \epsilon_{i,i+2} k_{i+2} )+\nonumber\\
%B k_i ( \epsilon_{i-2,i-2} - \epsilon_{i-1,i-1} )
%\label{eq:stress}
%\end{eqnarray}
%We apply the following strain:
%\begin{equation}
%\mathbf{\varepsilon}=\left(
%\begin{array}{ccc}
%\epsilon_1& \epsilon    & 0\\
%\epsilon  & \epsilon_1  & 0\\
%0 & 0   & \epsilon_2 
%\end{array}\right)\;
%\label{eq:def3}
%\end{equation}
%where $\epsilon_2=-2\epsilon_1$. Then
\begin{equation}
\left\{ \begin{array}{rl}
&\Omega^x_{stress} = C\epsilon k_y + B \Delta\epsilon k_x
\\
\\
&\Omega^y_{stress} = -\left[C\epsilon k_x + B \Delta\epsilon k_y\right]
\\
\\
&\Omega^z_{stress} = 0
\end{array} \right.
\label{eq:1}
\end{equation}
where $\Delta\epsilon=(\epsilon_1-\epsilon_2)$.
To facilitate the discussion let's write also explicitly
the components of the Dresselhaus field:
\begin{equation}
\left\{ \begin{array}{rl}
&\Omega^x_{D} = 2\gamma k_x\left(k^2_y-k^2_z\right)
\\
\\
&\Omega^y_{D} = 2\gamma k_y\left(k^2_z-k^2_x\right)
\\
\\
&\Omega^z_{D} = 2\gamma k_z\left(k^2_x-k^2_y\right)
\end{array} \right.
\label{eq:2}
\end{equation}   
Then we can write
\begin{equation}
\overline{\Omega^2_x}=
\overline{(\Omega^x_{stress})^2}+\overline{(\Omega^x_{D})^2}+2\overline{\Omega^x_{stress}\Omega^x_{D}}
\end{equation}
The first integral on the RHS is
\begin{eqnarray}
&&\overline{(\Omega^x_{stress})^2}=\left(C\epsilon\right)^2\overline{k^2_y}+\left(B\Delta\epsilon\right)^2\overline{k^2_x}+
\left(2 C B \epsilon \Delta\epsilon\right)\overline{k_x k_y}=\nonumber\\
\nonumber\\
&&\frac{1}{3}\left[\left(C\epsilon\right)+\left(B\Delta\epsilon\right)^2\right]k^2
\end{eqnarray}
The second integral on the RHS is
\begin{eqnarray}
\overline{(\Omega^x_{D})^2}=4\gamma^2\overline{k^2_x\left(k^2_y-k^2_z\right)^2}=\gamma^2 k^6 \frac{16}{105}
\end{eqnarray}
The third integral on the RHS is
\begin{eqnarray}
&&\overline{\Omega^x_{stress}\Omega^x_{D}}=\left(C\epsilon \gamma\right)\overline{k_x k_y\left(k^2_y-k^2_z\right)}+\nonumber\\
&&+\left(B\Delta\epsilon\gamma\right)\overline{k^2_x\left(k^2_y-k^2_z\right)}\nonumber=0
\end{eqnarray}
So we get
\begin{equation}
\overline{\Omega^2_x}=\frac{1}{3}\left[\left(C\epsilon\right)^2+\left(B\Delta \epsilon\right)^2\right]k^2
+\gamma^2 k^6 \frac{16}{105}
\end{equation}
In a similar way we obtain $\overline{\Omega^2_y}=\overline{\Omega^2_x}$
%\begin{equation}
%\overline{\Omega^2_y}=\frac{1}{3}\left[\left(C\epsilon\right)^2+\left(B\Delta \epsilon\right)^2\right]k^2
%+\gamma^2 k^6 \frac{16}{105}
%\end{equation}
and
\begin{equation}
\overline{\Omega^2_z}=\gamma^2 k^6 \frac{16}{105}
\end{equation}
For the off-diagonal components we get
\begin{eqnarray}
&&\overline{\Omega_x\Omega_y}=\overline{\Omega_y\Omega_x}=\overline{\Omega^x_D\Omega^y_D}+
\overline{\Omega^x_D\Omega^y_{stress}}+\nonumber\\
&&+\overline{\Omega^x_{stress}\Omega^y_D}+
\overline{\Omega^x_{stress}\Omega^y_{stress}}
\end{eqnarray}
The first three integrals on the RHS are equal to zero.
With the form of strain field given by Eq. (23) the 
last term can be written as
\begin{eqnarray}
&&\overline{\Omega^x_{stress}\Omega^y_{stress}}=\nonumber\\
&&-(C\epsilon)^2\overline{k_y k_x}+
(B\Delta\epsilon)^2\overline{k_x k_y}
-CB\epsilon\Delta\epsilon\overline{(k^2_x+k^2_y)}=\nonumber\\
&&-\frac{2}{3}CB\epsilon\Delta\epsilon k^2
\end{eqnarray}
All other $\overline{\Omega_i\Omega_j}$ are equal to zero.
Therefore according to equations (1) and (2)
for the spin relaxation time of an electron with energy E we find
\begin{widetext}
\begin{equation}
\frac{1}{\tau_{x,x}(k)}=\frac{1}{\tau_{y,y}(k)}=\gamma_l\tau_p(E)\left[\frac{1}{3}\left[\left(C\epsilon\right)^2+\left(B\Delta \epsilon\right)^2\right]k^2
+\gamma^2 k^6 \frac{32}{105}\right]
\end{equation}
\begin{equation}
\frac{1}{\tau_{z,z}(k)}=\gamma_l\tau_p(E)\left[\frac{2}{3}\left[\left(C\epsilon\right)^2+\left(B\Delta \epsilon\right)^2\right]k^2
+\gamma^2 k^6 \frac{32}{105}\right]
\end{equation}
\begin{equation}
\frac{1}{\tau_{x,y}(k)}=\frac{1}{\tau_{y,x}(k)}=
-\gamma_l\tau_p(E)\frac{2}{3}\left( B C \epsilon \Delta\epsilon \right)k^2
\end{equation}
\end{widetext}

Then the average spin relaxation time is
\begin{widetext}
\begin{equation}
\frac{1}{\tau_{x,x}}=\frac{1}{\tau_{y,y}}=\gamma_l\left[\frac{1}{3}\left[\left(C\epsilon\right)^2+\left(B\Delta \epsilon\right)^2\right]\frac{m^2}{\hbar^2}\langle v^2 \tau_p(E)\rangle
+\frac{32}{105}\gamma^2\frac{m^6}{\hbar^6}\langle v^6 \tau_p(E)\rangle\right]
\end{equation}
\begin{equation}
\frac{1}{\tau_{z,z}}=\gamma_l\left[\frac{2}{3}\left[\left(C\epsilon\right)^2+\left(B\Delta \epsilon\right)^2\right]\frac{m^2}{\hbar^2}\langle v^2 \tau_p(E)\rangle
+\frac{32}{105}\gamma^2\frac{m^6}{\hbar^6}\langle v^6 \tau_p(E)\rangle\right]
\end{equation}
\begin{equation}
\frac{1}{\tau_{x,y}}=\frac{1}{\tau_{y,x}}=-\frac{2}{3}\gamma_l B C \epsilon \Delta\epsilon \frac{m^2}{\hbar^2}\langle v^2 \tau_p(E)\rangle
\end{equation}
\end{widetext}
where $v=\hbar k/m$ and the brackets $\langle~~\rangle$ denote averaging over energies. For example for 
the Maxwell distribution
$\langle v^{2r} \tau_p(E)\rangle = \left(\frac{k_BT}{m}\right)^r\left(2r+1\right)!!\tau_p$.

By transforming the above tensor to the principal system of coordinates we obtain:
\begin{widetext}
\begin{equation}
\frac{1}{\tau_{\perp,1}}=\gamma_l\left[\left[\frac{1}{3}\left(C\epsilon\right)^2+\left(B\Delta \epsilon\right)^2\right]\frac{m^2}{\hbar^2}\langle v^2 \tau_p(E)\rangle
+\frac{32}{105}\gamma^2\frac{m^6}{\hbar^6}\langle v^6 \tau_p(E)\rangle\right]+\frac{2}{3}\gamma_l B C \epsilon \Delta\epsilon \frac{m^2}{\hbar^2}\langle v^2 \tau_p(E)\rangle
\label{eq:inplane1}
\end{equation}
\begin{equation}
\frac{1}{\tau_{\perp,2}}=\gamma_l\left[\frac{1}{3}\left[\left(C\epsilon\right)^2+\left(B\Delta \epsilon\right)^2\right]\frac{m^2}{\hbar^2}\langle v^2 \tau_p(E)\rangle
+\frac{32}{105}\gamma^2\frac{m^6}{\hbar^6}\langle v^6 \tau_p(E)\rangle\right]-\frac{2}{3}\gamma_l B C \epsilon \Delta\epsilon \frac{m^2}{\hbar^2}\langle v^2 \tau_p(E)\rangle
\label{eq:inplane2}
\end{equation}
\begin{equation}
\frac{1}{\tau_{\parallel}}=\gamma_l\left[\frac{2}{3}\left[\left(C\epsilon\right)^2+\left(B\Delta \epsilon\right)^2\right]\frac{m^2}{\hbar^2}\langle v^2 \tau_p(E)\rangle
+\frac{32}{105}\gamma^2\frac{m^6}{\hbar^6}\langle v^6 \tau_p(E)\rangle\right]
\end{equation}
\end{widetext}
Where $\parallel$ denotes axis parallel to the vector ${\bf N}=(0,0,1)$ and 
$\perp$ perpendicular to it. Namely, $(\perp,1)$ and $(\perp,2)$ are the axes along the [1$\bar{1}$0] and 
[110] crystal directions, respectively.
Equations (\ref{eq:inplane1}) and (\ref{eq:inplane2}) signal an in-plane
anisotropy induced by the simultaneous presence of $\epsilon$
and $\Delta\epsilon$.

\bibliography{cgwzb-gas}% Produces the bibliography via BibTeX.

\end{document}